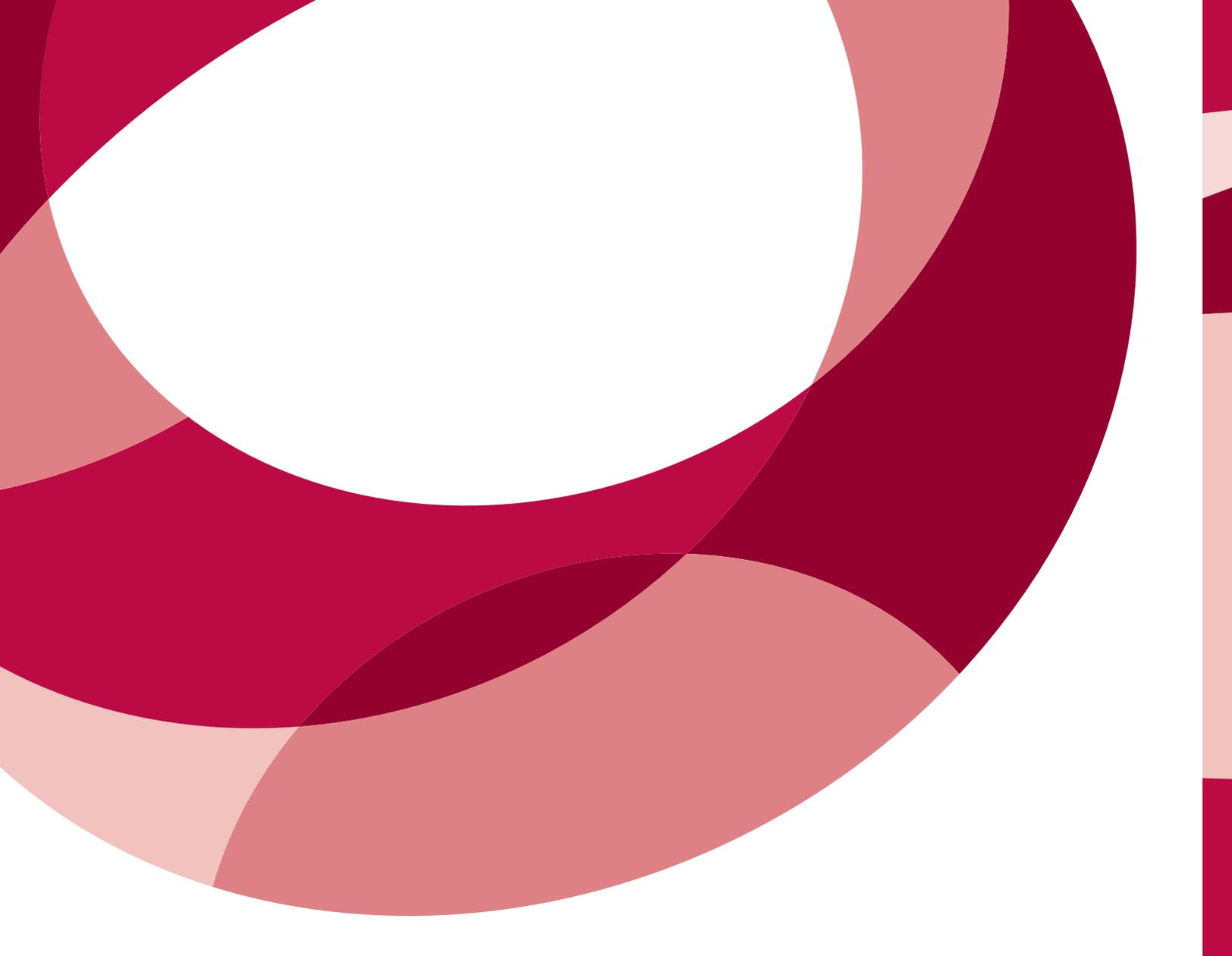

# Community Driven Approaches to Research in Technology & Society CCC Workshop Report

March 2024

The material is based upon work supported by the National Science Foundation under Grant No. 1734706. Any opinions, findings, and conclusions or recommendations expressed in this material are those of the authors and do not necessarily reflect the views of the National Science Foundation.

# Community Driven Approaches to Research in Technology & Society CCC Workshop Report

## March 2024

**Workshop Organizers**

Suresh Venkatasubramanian (Brown University)*
Timnit Gebru (Distributed Artificial Intelligence Research Institute)*
Ufuk Topcu (University of Texas at Austin)*

**With support from:**

Haley Griffin (CCC)*
Leah Rosenbloom (Brown University)*
Nasim Sonboli (Brown University)*
Cat Gill (CCC)
Mary Lou Maher (CCC)
Ann Schwartz (CCC)

*Report authors

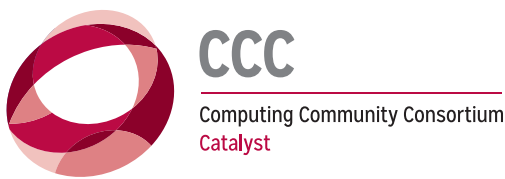

Supported by
MacArthur
Foundation
www.macfound.org

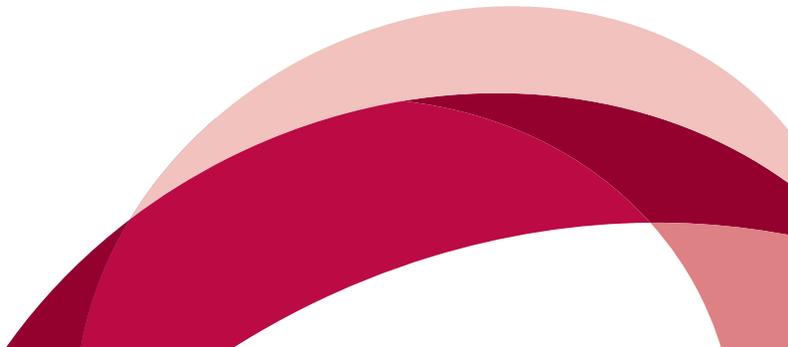

# Executive Summary

The Community Driven Approaches to Research in Technology & Society workshop on community based research in computing, sponsored by the CCC and the MacArthur Foundation, was held on May 8 and 9 2023, bringing together 53 people, roughly half are scholars in computing at universities, nonprofits and industry, and half are advocates of communities and members of communities whose lives are severely impacted by the use of AI systems (see Appendix for Workshop Participants list). Many people had intersecting identities belonging to subsets of these groups.

**Workshop Goals:** The goals of the workshop were to bring together researchers in computing and people who have intimate knowledge of the impacts of AI systems either through their lived experiences or work on advocacy to (1) build a coalition comprising community partners and academic researchers studying the societal impact of automated decision making in society, (2) identify key research questions and directions for progress along the areas identified in the Blueprint for an AI Bill of Rights, and (3) establish partnerships between advocates and researchers to push forward the principles and expectations articulated in the workshop.

**Workshop Activities:** To achieve the goals of the workshop we organized a series of level-setting talks by computing researchers and members of advocacy groups talking about the main issues they face pertaining to AI systems, the manner in which computing research can support their needs, best practices for collaboration, and the ways in which research has supported their efforts thus far. We also organized breakout sessions organized around these themes, and synthesized the discussions to come up with actionable next steps as well as recommendations for the three workshop goals outlined above.

**Research Directions:** Based on our workshop activities, we outlined three ways in which research can support community needs: (1) Mapping the ecosystem of both the players and ecosystem and harm landscapes, (2) Counter-Programming, which entails using the same surveillance tools that communities are subjected to observe the entities doing the surveilling, effectively protecting people from surveillance, and conducting ethical data collection to measure the impact of these technologies, and (3) Engaging in positive visions and tools for empowerment so that technology can bring good instead of harm.

**Mechanisms for Effective Collaboration:** In order to effectively collaborate on the aforementioned directions, we outlined seven important mechanisms for effective collaboration: (1) Never expect free labor of community members, (2) Ensure goals are aligned between all collaborators, (3) Elevate community members to leadership positions, (4) Understand no group is a monolith, (5) Establish a common language, (6) Discuss organization roles and goals of the project transparently from the start, and (7) Enable a recourse for harm.

**Recommendations:** We recommend that anyone engaging in community-based research (1) starts with community-defined solutions, (2) provides alternatives to digital services/information collecting mechanisms, (3) prohibits harmful automated systems, (4) transparently states any systems impact, (5) minimizes and protects data, (6) proactively demonstrates a system is safe and beneficial prior to deployment, and (7) provides resources directly to community partners.

We recommend that funding agencies who support academic research that is relevant to communities (1) solicit community input when designing solicitations and constructing review panels, (2) include additional proposal evaluation metrics to address computing participation and leadership, (3) require impact reports if communities are affected, (4) account for the additional amount of time required to build relationships with community partners, and (5) streamline funding infrastructure for community partners.

We recommend that academic institutions (1) factor community impact into academic evaluation processes, and (2) provide funding for community-based research projects.

We recommend that individual researchers (1) follow our listed norms for community based research, (2) advocate for the community when given a leadership role, (3) expand the vision of what is considered a valuable output to account for positive community work, (4) understand the tension between generalized insights, that academics are oftentimes aiming for, and underlying problems and solutions deeply rooted in humans and communities, that are largely heterogeneous and context-specific, and (5) generate research artifacts that protect community data, which oftentimes requires them to have limited access.

# 1. Introduction

## 1.1. Background

Artificial intelligence (AI) reaches into every aspect of our lives today with many intended and unintended consequences for people and communities. The focus of this workshop was to enable conversations between computing researchers and people who have experienced the full range of effects of AI systems to better understand the impacts of technology within communities, the current state of research that seeks to improve lives in an algorithm-driven world, and the key obstacles preventing this work from coming to fruition.

The White House Office of Science and Technology released the Blueprint for an AI Bill of Rights[i], a document that lays out key protections that people and communities are entitled to in an AI-powered world. The Blueprint notes that "AI and other data-driven automated systems most directly collect data on, make inferences about, and may cause harm to individuals. But the overall magnitude of their impacts may be most readily visible at the level of communities." It continues by asserting that "the harms of automated systems should be evaluated, protected against, and redressed at both the individual and community levels."

## 1.2. Workshop Goals

The workshop participants noted that the Blueprint's focus on community harms provided a starting point for a broader discussion of the concerns of community-centered research on the impact of automated decision making in society.

The goals of the workshop were:

◗ to build a coalition comprising community partners and academic researchers studying the societal impact of automated decision making in society;

◗ to identify key research questions and directions for progress in addressing those impacts;

◗ to establish partnerships between advocates and researchers to push forward the principles and expectations articulated in the workshop.

## 1.3. A Note on the Term "Community"

While both the workshop title and this report make many references to the word "community," workshop participants pointed out that the term itself is ambiguous – anyone can be part of any community, and each person can also be part of multiple communities. In this workshop, we primarily use this term to refer to people and groups who bear the brunt of the negative impacts of AI systems, rather than those shaping its production and governance. While computing researchers can also be part of this group, they usually have better access to resources, and more power to shape research priorities in the field of AI. Thus, when we use the term "community," for instance in phrases like community-based, community-driven, community-informed, or community-centered, we are referring to those affected by AI systems.

## 1.4. On How to Read this Report

This workshop was initially conceived of and constructed as a conversation between computing researchers and community participants impacted by technology, and specifically by automated decision-making. It became clear early on in the workshop that one of the most important outcomes would be to elevate and communicate the challenges and concerns our community partners face when participating in research work that sought to address the impact of technology. In response to these challenges and concerns, this report articulates models of how to conduct this research in ways that are

---

community-driven, and recommendations for how different institutional actors (researchers, universities, and research funding agencies) could incorporate this feedback into their practices and incentive structures. In doing so, this report seeks to rebalance the power dynamics between researchers and their community partners by centering the concerns of communities first, and giving them voice. The critique of current research practices that this report documents is intended to help steer conversations and actions towards research work that benefits all those who are currently impacted.

## 1.5. How We Assembled

In general, ideas for Computing Community Consortium (CCC) visioning workshops are generated from CCC Council Members, federal agencies, and/or directly from the computing community as responses to an open call for visioning proposals. For this workshop, organizers reached out to CCC leadership with a proposal for a workshop that responds to the Blueprint for the Artificial Intelligence Bill of Rights. During his time at the White House Office of Science and Technology Policy (OSTP), Suresh Venkatasubramanian, a former CCC Council member and a Professor at Brown University, was part of a team that created the Blueprint. The Blueprint development process included collecting input from a diverse group of stakeholders, including members of communities affected by AI systems, researchers and practitioners, and policymakers. Another workshop organizer, Timnit Gebru, runs a research institute that is focused on performing community-based AI research. The final workshop organizer was CCC council member Ufuk Topcu, who works on responsible AI.

As we began inviting individuals to the workshop, we aimed for as broad of a participatory body as possible, with an approximate 50/50 split of computing researchers and community members. We also invited civil society organizations who advocate for communities. Our workshop had 53 attendees, representing 43 unique institutions/organizations. We also strived for broad participation across many other metrics of diversity, including race, gender, disability, class, sexual orientation, geographical location, institutional affiliation, and research area. See the Appendix for the Workshop Participants list.

## 1.6. Workshop Activities, Process, and Mechanics

The workshop organizers established Chatham House rule[2] at the beginning of the workshop. Discussions that occurred at the workshop, both in breakout sessions and the main sessions, and including participants' lived experiences, inform this report. However, no remarks will be attributed to specific participants in accordance with the rule.

Day 1 began with a talk about the goals of the workshop, expectations for participants, context on how the workshop came to be, and the agenda for the day. A presentation was made about the process of creating the Blueprint for an AI Bill of Rights, and its potential impact. A community organizer spoke about the research they and their team are doing to map out the types of technologies that are being used to combat illegal immigration in the US, and what this research taught them about the harms the technology can cause the immigrant community and how that community might protect itself. Another speaker discussed the types of issues that disabled people face while using AI systems, ranging from biased image captions, to applications that can cause severe harm, or even death, if they have high false positive rates. They outlined a number of opportunities for research directions that can make AI systems more accessible and safer to use for disabled people.

Subsequently, workshop participants were separated out into smaller groups for two breakout sessions. Each breakout session (which also includes a third breakout on Day 2) had a theme, and three questions for discussion. After each breakout, all groups returned to the main workshop area to share their insights. Participants were assigned to different groups for each breakout session.

The first breakout session was organized around the topic: "Community-Driven Research: existing problems, solutions, and wish lists". The three discussion questions were,

---

[2] https://www.chathamhouse.org/about-us/chatham-house-rule

◗ **S1-Q1: Problems.** In your experience, what is a significant problem you have faced with technology that affects our civil rights and liberties, opportunities for advancement, and/or access to critical services?

◗ **S1-Q2: Existing solutions**. In your work, what kinds of solutions/frameworks have you seen to be effective in ensuring that community concerns and needs are accounted for?

◗ **S1-Q3: Wish list.** What are three (max) of the most promising/effective ways to connect people and technology that you wish would exist to solve the problems mentioned earlier?

The second breakout session was organized around the topic: "Best practices for multi-stakeholder collaborations." The following prompt was provided to each group.

"In the previous session, we focused on the technologies that currently impact people in many ways, and how we would like them to change/evolve. To do this, it's clear we need conversations among stakeholders with different backgrounds, lived experiences, and expertise. *And we need to identify key ingredients that make these conversations successful.*"

The three discussion questions were,

◗ **S2-Q1: Positive outcomes.** What are good experiences you've had collaborating with stakeholders that led to a positive outcome for all involved?

◗ **S2-Q2: Pitfalls.** What are the ways in which things went wrong?

◗ **S2-Q3: Key ingredients.** Identify 3-5 key ingredients for successful collaboration as well as ways to uplift each of these practices you identify.

Day 2 began with a "Reflections of Day 1" session led by the workshop organizers. They synthesized the overarching themes from Day 1, and introduced the topics for Day 2.

Two researchers and practitioners who also come from communities often negatively impacted by AI discussed successful models of community-driven research. One speaker presented their organization's work on enabling a sovereign digital future for indigenous people, and their team's process of building language technology for language revitalization that prioritized the desires of their community. The second speaker discussed the use of speculative design practices to work with Black communities, especially older Black adults, to design technologies that are useful and culturally relevant to them.

The third breakout session was organized under the theme: "Opportunities for deep engagement for communities". The discussion questions were,

◗ **S3-Q1: Research opportunities.** What are open research questions (based on the archetypes we discussed in the synthesis) that, if addressed, would make a difference in your work?

◗ **S3-Q2: Mechanisms for collaboration.** What is your currency in your work? What are new kinds of currency we need to be building?

◗ **S3-Q3: Community roles.** In your mind, how would you describe the different ways that communities can and should be engaged?

The final session was a full-group synthesis of next steps. We did a whole-group brainstorming session on:

◗ **S4-Q1: Needs.** The needs that each individual's community is facing that could be solved using technology/requests for technology creation or implementation that does or doesn't currently exist.

◗ **S4-Q2: Solutions/skills.** The solutions/skills that people in the room knew of or could envision that have the potential to help solve the problems described in part S4-Q1.

The remainder of this report is organized as follows. Section 2 provides important context by synthesizing and sharing the perspectives and experiences of community partners who conduct research, and work with academic partners. Section 3 identifies models for successful research directions, and Section 4 provides examples of effective mechanisms of collaboration (as well as mechanisms to avoid). Section 5 provides specific recommendations for researchers, institutions, and research funding agencies.

## 2. Concerns Voiced by Workshop Participants

One of the goals of the workshop was to identify the types of issues that arise when doing community-centered research. The workshop participants discussed a number of issues that largely fall into the five themes below. Many of these problems reflect structural dynamics that are common in community-centered research, even if they are beyond the scope for any individual to address. To that end, the report provides (forward) references to recommendations for structural change where appropriate.

### 2.1. Incentive Structures for Researchers can Lead to Exploitation

The most frequently brought up issue was the fact that the incentive structures that motivate academic researchers can lead to exploitation during collaborations with communities. This can take a number of forms. First, many organizers and others at the front lines that advocate for community rights may not have the credentials that are rewarded by funding agencies and academic institutions. This can create a power dynamic where those who are eligible for funding are awarded resources, rather than community members with more knowledge and lived experience about the issues they need to address. Thus, once a collaboration is established, there can be a power dynamic where community members are beholden to the goals of the funded research proposal and are not compensated, or treated as equal collaborators (Recommendation 4.2. & 4.3.). To address this, workshop participants encourage funders to provide resources directly to community collaborators (Recommendations 4.1. & 5.1.g.). Researchers noted that funding agencies, like the National Science Foundation (NSF), have infrastructure to fund students, but that there are more restrictions to fund community based collaborators who may not be students or academic researchers (Recommendations 4.1. & 5.2.e.).

This type of power dynamic also complicates the ethics of compensation, especially if someone feels compelled to participate because they need the money. For instance, if people are offered compensation to provide their data, it may not actually be a choice for those who are deciding between obtaining funds to feed their families versus providing their data. Recent news articles have noted that oftentimes funding for nonprofits are not earmarked for essential things like security and data protection[3]. Our community participants commented that while community-based research often requires data collection, they are concerned about how that data is secured (Recommendation 5.1.e.). One participant noted that some data on education technology platforms used by schools is not required to be encrypted. Even systems which are encrypted encourage mass, centralized collections of sensitive data which can and have been breached by cyberattacks[4].

This research can also be problematic if information on how to be involved is only disseminated digitally. Community members discussed examples of when the lack of access or use of digital technology caused harm. For instance, during the COVID-19 pandemic some older Black adults that do not have internet access were not able to access information about vaccinations because it was only shared digitally. One of the workshop participants also gave an example of a deaf woman who was told that she would only be able to receive updates on her place in a kidney transplant list over the phone (Recommendation 5.1.b.).

Workshop participants also noted a changing power dynamic across the duration of the project, with community based collaborators being viewed as more valued at project set up time versus during the course of the project itself (Recommendation 4.6.). Some noted that they cannot pull out of the project once this happens, and the lack of adequate whistleblowing or safety mechanisms for people to report such instances (Recommendation 4.7.).

### 2.2. Tokenization of Community Representatives

Another issue that was raised by workshop participants was the common occurrence of tokenizing community representatives: that is, certain people who are part of a particular demographic of interest are selected often to represent the entire demographic group, and repeatedly tapped to be the voice of the community. This can be detrimental for a number of reasons. First, this small set of people may not represent the interests of the community, but may in fact have views that are more aligned with the goals and expectations of the researchers. Secondly, no community is a monolith and as such, a couple of people and organizations cannot be representative of entire groups of people (Recommendation 4.4.). Workshop participants brought up the question of "what does community mean?" There are many different types of communities and each member plays different roles depending on the context (Recommendation 5.1.a.).

---

[3] https://www.crainsdetroit.com/nonprofit/nonprofits-low-it-investment-mixed-staffing-models-are-easy-targe ts-cyber-attacks
[4] https://www.nytimes.com/2022/07/31/business/student-privacy-illuminate-hack.html

In addition to tokenism, the community expressed concern about lack of partnerships in their participation in research. One participant described a situation in which their organization was listed in a proposal as a collaborator when the researchers had not even reached out to them about working together. This can happen when researchers know that they are required to check the box of having a community based partner, but lack the incentives to have an equal partnership. Many community members also noted their experiences with being constantly tapped by researchers for input without compensation, while their input was ignored, their guidance was not implemented, and there was no follow through from the researchers (Recommendation 4.6. & 5.4.e.).

## 2.3. The Types of Research that "Deserve" Funding

The final theme that workshop participants brought up was that the very notion of what constitutes research, what questions are important, and what activities are worth funding can end up weeding out important community based work. One participant expressed the concern that the social currency for academics is grants, teaching, publications, etc., and that community work is not as valued (Recommendations 4.2. & 4.6.). This is reflected in the funding agencies' allowable expenses not recognizing community liaisons who are not students or academics (Recommendation 5.2.e.). They noted that solicitations should require more substantial community collaboration sections beyond broader impact statements for community-based research. People also noted that on the occasion that there are research calls pertaining to community concerns, they are often about identifying the harms a community is facing. However, communities many times know the harms they are facing themselves, and do not need resources to identify them, but rather to address them (Recommendation 5.1.a.). That is, funding should be directed at addressing harms known to community members by involving them in the research, rather than on discovery of already-known harms.

In addition, the solutions that are considered fundable may be less effective in the long term than what community-based grassroots organizations would propose (Recommendation 5.2.a.). For instance, one participant noted that when it comes to social media platforms, "much of the conversation was first about how to better accommodate some of our marginalized communities, such as using their own language on Facebook". Yet, there are many people in their communities that believe in reducing the use of some types of social media platforms, noting that the community does not want "imperialist colonialist platforms". Instead, the participants discussed the types of works that are funded include those related to diversifying various languages on existing platforms rather than having alternatives to these centralized systems. These types of differences in beliefs as to what is best for a particular community can also create turmoil within the community, which highlights the importance of having more community participation to represent such diverse groups.

## 2.4. The Agenda-Setting Role of Industry and Government Actors and their Influence on Research

A theme workshop participants returned to repeatedly was the role of industry and government actors in setting agendas, framing problems, and influencing (via funding) the ways in which academic researchers approach research questions that have societal impact. While a full examination of the role of industry and government actors was beyond the scope of this workshop[5], there were a few important concerns relevant to academic researchers that we summarize here.

---

[5] In particular, workshop organizers explicitly avoided inviting participants from industry and government so as to allow for a free and open discussion.

A first concern was the overrepresentation of corporate perspectives while discussing any type of research related to technology, and the lack of representation from civil society (Recommendation 4.3.). This results in a number of issues. This overrepresentation leads to technology-first, rather than community-first approaches to problem solving (Recommendation 5.1.a.). For instance, instead of asking whether a specific product should exist in the first place, its application for one socially beneficial purpose can be used as justification for its existence even if it mostly causes harm. The tokenization of individuals in a community (Recommendation 4.4.) discussed above has often been used to justify this. For example, Chancey Fleet discusses how the needs of blind people have been used as justification for wide-spread use of face recognition based systems[6] without any blind people being consulted in the process. She notes that "Disabled people value privacy and our communities' freedom from surveillance as much as anyone" and that they should not be made a justification to increase surveillance. Related to this is the way in which corporate funding (much like government funding described above) can steer research agendas and what are considered valuable problems and solutions (Recommendation 5.1.g.)

A second concern was that many of the systems used to surveil and impact communities are developed by government entities, who in turn are also the ones funding and shaping academic research (Recommendation 5.1.f.). This makes it difficult for academics to critique these systems or even reframe the problems being addressed (Recommendation 5.1.d.). For example, many essential government programs, such as welfare, are being automated[7] (Recommendation 5.1.b.), and while academics have participated in the design and evaluation of these systems, it is more challenging to critique the way in which these systems are being used to increase surveillance and control of already marginalized communities (Recommendations 5.1.c. & 5.1.d.).

All of the issues mentioned in this section point to the need to incentivize research processes that support the "repeated quiet work" that often characterized engaged community work (Recommendation 5.2.d.). In the next section, we discuss research themes that emerged from our workshop and the ways in which they can address the needs of people at the margins.

# 3. Models for Effective Research

Workshop participants provided many examples of effective research projects as well as ideas for future collaborations. The research directions mainly fell into the themes below.

## 3.1. Mapping Projects

One challenge raised by workshop participants is understanding the broader context in which a sociotechnical system is deployed, and how it will impact people. There are two important research approaches that support this understanding.

### 3.1.a. Players and Ecosystem

One workshop participant described research that they conducted to identify and map the different organizations involved in surveilling immigrants in the US, and the relationships between them. These include government agencies and officials, corporations providing software and services, key databases that drive investigations and other actions, as well as underlying laws, regulations, and policies. Performing this research helped the participant and their organization understand the discrepancy between stated policies and the activities being conducted on the ground, which is often mediated by opaque technologies.

### 3.1.b. Harm Landscapes

The ways in which communities experience harm from technology deployment can vary greatly depending on the group, technology, and sector (health, policing, housing, etc.). While broad discourse around technology harms focuses on high level and cross-cutting concerns, mitigation strategies require a much more detailed understanding of how harms manifest within a specific context.

The key research questions here are identifying harms in a specific context, and laying out the *harm landscape*. This requires community-led research, because it is only at the point of impact that we can understand the ways in which technology-mediated harms emerge (Recommendation 5.1.a). It also requires us to recognize that such research is valid and valuable in and of itself, even without accompanying mitigation strategies. This point is important to reiterate – in recent years, while we have seen more research that maps out harm landscapes, there has been consistent pushback under the premise that harm identification without providing corresponding solutions is not useful (Recommendation 5.4.c).

## 3.2. Counter-Programming

Technology tools, especially those based on data collection and machine learning, are used to measure and predict behavior, and are a common source of harm. One way to "flip the script" is to recognize that these same tools can be used by those being subject to them, to monitor and observe the (more powerful) entities subjecting communities to these tools.

Workshop participants advocated for the use of technology for "counter-programming" in this manner. They identified the following three different ways in which technologies could be repurposed.

### 3.2.a. Tools to Watch the Watchers

Communities have long used surveillance technology to "surveil the surveillers". For example, cell phone cameras have been used to provide real time evidence of unarmed Black people being killed by police–starting world wide movements like the Black Lives Matter movement. Most recently, the Algorithmic Justice League has mobilized travelers to report (via a simple web form[8]) on the use of facial recognition systems in airports and whether the deployment by the TSA lives up to its promise of being opt-in and voluntary. Communities need tools that help them monitor those who seek to observe and surveil them, so that they can provide counter-narratives and bring to light problematic practices they endure.

### 3.2.b. Protecting People and Communities from Surveillance

A predominant narrative of harm associated with technology is in the way it is used to surveil and oppress communities. Technology itself however can be repurposed to instead serve communities and protect them from these harms. Such a *reimagining* of technology has been called out in the context of cryptography and computer security[9], and represents a different way of approaching technology design that seeks to invert power structures rather than support them.

While online tools for social interaction are a powerful way to mobilize communities and support activism, they are vulnerable to digital surveillance and are not secure enough to be used in situations where activists are at risk of being placed under scrutiny. Indeed, Signal[10] is one of the only applications currently reliable and secure enough for online communication and is itself under threat from proposed surveillance legislation. One ongoing project described by workshop participants seeks to use cryptographic protocols to facilitate activists' work in a privacy-preserving way by translating existing physical trust-building protocols into digital space.

Another example of reimagining the use of technology is the Glaze[11] system developed at the University of Chicago. Artists have become increasingly concerned with the way generative AI systems are absorbing their content during training and repurposing it without any credit, acknowledgement, or compensation. Glaze is a tool that allows artists to modify their digital images in subtle ways that do not impair the visual effect, but render the images impossible to ingest and reproduce in the style of specific artists by image generation deep learning systems such as Stable Diffusion[12] and Midjourney[13]. Another example

---

is the new tool Nightshade[14] that subtly modifies artist-generated images so that if they are used without consent as training data for generative AI, they poison the resulting model.

### 3.2.c. Effective Measurement

Measurement and data collection are powerful tools to aid advocacy work. Regular measurements of outcomes demonstrate the ways in which communities are impacted by technology, and also establish baselines and trends to determine whether advocacy is having an effect, and how much. For example, activists have long advocated for documentation and measurement of where and to what extent facial recognition systems are being used to investigate and detain individuals. Measurements can also create new kinds of currency and new ways to support communities in their advocacy. For example, if communities are able to quantify the impact (say) of housing appraisal algorithms by documenting and measuring home valuations, it provides a powerful way to illustrate where such tools might benefit homeowners and where they might harm them.

The research questions here involve how communities can effectively and easily do the kinds of measurement and data collection that can help them advocate for themselves, and what kinds of tools can be built to help them in this effort.

### 3.3. Positive Visions and Tools for Empowerment

Technology can, and should, bring joy rather than be a source of harm, or something to defend against. Workshop participants emphasized the importance of positive visions of the future, and a focus on design that empowers and enables. Speakers described participatory design research in which older Black adults kept diaries of their interactions with digital home assistants, and in doing so identified ways in which they were being forced to adapt to the technology – for example when asking about illness and medication – rather than have it be useful for them[15]. Participants described community driven design workshops in which participants imagined future ethical uses of technology, and used physical artifacts to dream of ways in which technology could be beneficial to them in the future[16]. Participants described the image captioning systems that Blind people use together with their screen readers. These systems are extremely valuable to those that need them, but are often error prone. Systematic investigations – audits – of these technologies can lead to improvements that are beneficial to those that need them the most.

The key research question here is "tools for empowerment". Such approaches do not require a deep understanding of technology, but rather a call for a structured design process that empowers people to dream of better tech futures.

## 4. Models for Effective Collaborations

In addition to outlining research themes that can support grassroots organizations, workshop participants outlined a number of key considerations for successful collaborations across different disciplines and institutions with varied incentive structures.

### 4.1. No Free Labor

Community members are sometimes expected to work on research projects without any compensation, even though the artifacts associated with the projects oftentimes do not help them advance in their careers or earn more money. In addition, the types of research that people do can further traumatize them or put them in harm's way. Successful collaborations do not expect free labor from community members, but rather appropriately compensate them with monetary funds and/or other resources. Community based collaborators should gain sufficient credit in the currency that is valued by them, and get access to reports and other information that can be leveraged in their organizing efforts.

---

[14] https://www.technologyreview.com/2023/10/23/1082189/data-poisoning-artists-fight-generative-ai/

[15] https://dl.acm.org/doi/10.1145/3491102.3501995

[16] https://www.tawannadillahunt.com/wp-content/uploads/2021/05/elicitingtechfutures.pdf

## 4.2. Goal Alignment

Participants noted that the goals of all the collaborators need to be aligned. They asked who makes up a stakeholder? If marginalized people are in the same project as those who are responsible for their marginalization, there will clearly be tension, and the environment will not be one that allows those who are marginalized to fully participate. However, researchers can use this dynamic to have a stamp on their work saying that people from a specific demographic were in the room. Being in the room is a necessary but not sufficient condition for community based research. Researchers should avoid what professor Keeanga-Yamahtta Taylor calls predatory inclusion[17]. These research relationships should be authentic, based on empathy, and not forced. And people should be able to leave the collaboration at any time, and be made aware of their options for doing so.

## 4.3. Leadership by Community Members

The best collaborations are ones where community members are treated as equal collaborators, as either co-PIs or whatever leadership position is applicable during the project, and co-design the research questions and processes at every step. Besides adequate compensation, one way to ensure the partnership is equitable is authorship. However, authorship may not always represent useful currency for community partners and in some cases could be actively harmful. Its value should be estimated in context.

## 4.4. No Group is a Monolith

A recurring theme in the workshop was the fact that no group is a monolith, and no one should be assumed to be speaking on behalf of an entire group. Thus, researchers should not use the input of a handful of people from a particular group as representing all possible perspectives of said group. Workshop participants noted that it is important to meet people where they are, and not impose one's preconceived notions of how they should react to a specific issue. Workshop participants also brought up the importance of collaborations that are not centered only around the US, and finding ways to figure out how tech is used in different cultural contexts.

## 4.5. Common Language

Workshop participants noted the importance of having common language and terminology across collaborations aiming to do community centered research. Different groups of people may describe the impacts of a particular technology differently. Sometimes, translational work discussing works from large institutions in ways that people who may be harmed by those institutions' works understand, can be useful. To establish and continue this common language, it may help to appoint specific people to the task of facilitating constructive dialog among the different members of the research team.

## 4.6. Clarity about Organization Roles

Workshop participants pointed out transparency of the goals of the project, the potential challenges, and expectations of the different parties, as another important component of a successful collaboration. It is important to have frequent occasions to give and receive feedback, set the right expectations and boundaries together from the beginning, and to have mutual understanding of project goals and what success looks like for everyone involved. Having a point person assigned to enable and sustain interactions, and maintain an open line of communication, may be helpful.

## 4.7. Recourse for Harm

Some workshop participants pointed out that they often do not have any recourse for repairing harm. As an example, one of the workshop participants explained an instance where their input was requested while they reviewed an academic paper pertaining to sex workers. The workshop participant spent a lot of unpaid labor crafting a thoughtful review that outlined the harm that the paper could perpetuate, and suggested changes. None of the suggested changes were incorporated, and the paper was published. The workshop participant had no recourse. Collaborations between academic researchers and community based collaborators need to

---

have explicit steps for recourse and repair if the terms of the collaboration are breached and/or harm is caused.

## 4.8. Example of Successful Collaboration

One participant gave an example of a collaboration that they saw as a successful model, discussing a program called the Academic Autistic Spectrum Partnership in Research and Education (AASPIRE)[18], which is a collaboration between Portland State University and many community and academic organizations like Autistic Self-Advocacy Network, and Oregon Health and Science University. A workshop participant noted that this was the only example they knew of in the disability advocacy space that is academically housed and an equal partnership between disabled and abled researchers, with the co-directors being an abled and disabled researcher. In addition, according to the participant the project:

◗ Provides stable and secure jobs for autistic people.

◗ Carries out research from autistic people's perspective, starting with the questions they want the research to answer.

◗ Has a structure where research participants have decision making power.

The workshop participant noted that there are clear limitations and questions they have about various aspects of the collaboration. But they still believed it to be a useful example of how academic researchers can structure and resource their collaborations to provide secure and stable financial opportunities.

# 5. Recommendations

## 5.1. Broad Recommendations on Ensuring Ethical Community-Based Research

These broad recommendations are relevant to obtaining ethical approval for researchers who seek technological solutions through empowering communities. While many of these recommendations could be seen to be addressed in current applications for ethical approval in human subjects research, these recommendations focus on the concerns of the broader community in community-based research.

### 5.1.a. Community-Defined Solutions

Communities should be resourced to define community solutions themselves. Too often researchers focus on community opinions only to define problems, and then develop solutions that do not substantially take community input and needs into account. Researchers should strive to obtain broad community feedback about both problems and solutions.

### 5.1.b. Right to Not to be Digital

There should be non-automated alternatives to deploying digital services and collecting digital information when doing community-based research on the impact of future technology. There are members of the community that are of interest in community-based research that do not have access to technology. Others might have the means to use technology, but choose not to for a myriad of different reasons. These subgroups of the community are not heard, and are potentially disadvantaged by the fact that they are not online.

### 5.1.c. Harmful Automated Systems Prohibition

Automated systems have potentially negative consequences can perpetuate bias and harm. There should be a process to continuously and thoroughly monitor automated systems to ensure that they are not causing harm, and a mechanism to enact prohibitions towards any that are.

### 5.1.d. Public Transparency Regarding Systems Impact

People have the right to understand how a system is affecting them, so that they can choose whether they want to interact with the system, or implement changes to it. People will not necessarily know what questions to ask about technology until they are sufficiently educated about its implications.

---

[18] https://aaspire.org/?p=home

### 5.1.e. Data Minimization and Protection

Researchers should only collect and store community and individual data that is essential to their project, and instill robust privacy-enhancing technologies.

### 5.1.f. Proactive Regulations

The creator of an automated system should be required to demonstrate to the extent possible that the system is safe and beneficial before deploying a product, rather than placing the burden of proving harm on users after the fact. Frameworks for demonstrating safety should be developed for new technology for which it is not possible to demonstrate safety in all contexts. The onus on safety, privacy, and fairness should be on the organizations and entities putting out these systems (i.e. privacy or fairness by design).

### 5.1.g. Direct Resources to Community Partners

Regardless of the funding source of an initiative, community partners should be fairly compensated for their time, including any preparation time, travel time, resources required, etc. The currency for each individual in a community is different, and researchers should build a connection with them to figure out what they value most. While it is likely monetary, sometimes other values like privacy and authorship should supplement or replace monetary incentives.

## 5.2. Recommendations for Funding Agencies

These recommendations are directed towards funding agencies (and in particular government funding agencies) who support academic research, as well as those soliciting funding for research that impacts or is otherwise relevant to communities. The purpose of these recommendations is to incentivize meaningful and substantive engagement and leadership of community partners in the conception, evaluation, funding, and execution of research that impacts their lives.

### 5.2.a. Community Input When Designing Solicitations and Constructing Review Panels

Funding agencies should actively seek out feedback from community partners when designing solicitations for research proposals that have societal impact. In doing so, agencies should make efforts to:

◗ Expand the range of research that is considered 'within scope'. In particular, research about the impact and deployment of computing technologies should be an option, in addition to research that focuses on improving specific technologies.

◗ Avoid academic jargon and use language and terminology that is accessible and legible to those communities who are impacted.

◗ Avoid tokenizing and unpaid participation from community partners.

Funding agencies should make concerted efforts to ensure that steering committees, working groups, and other formal committees that assist in the development of solicitations, have representation and/or feedback from community members.

Solicitations for research that impacts communities should strongly encourage or require that community members be co-investigators on proposals, and regard their expertise as comparable to the credentials of academic researchers. Moreover, funding agencies should design mechanisms to allow community partners to lead/co-lead proposals.

Funding agencies regularly seek out different forms of diversity and specific expertise in review panels, including diversity in institution, level of seniority, and research credentials. Similarly, funding agencies should seek out community expertise when creating review panels.

### 5.2.b. During Proposal Review

Funding agencies should include additional proposal evaluation metrics that carry weight similar to traditional measures (such as intellectual merit and broader impact) that are more directly focused on community participation and leadership in research design and impact. These metrics should not just incentivize community engagement as part of the research, but also as part of the proposal design process, with demonstrated evidence of community input being taken into consideration.

### 5.2.c. During the Evaluation of Research Subsequent to Funding

Funding agencies should require impact reports if communities will be impacted by the research being developed.

### 5.2.d. Duration of Funding for Community-based Research

Doing meaningful work that is co-directed by community partners takes time to build relationships and understand the real problems at play. Funding agencies should seek to incentivize the longer-term, hard and slow work that builds relationships and design metrics for evaluation appropriately, rather than solely relying on metrics like publications and presentations.

### 5.2.e. Streamlined Funding Infrastructure for Community Partners

It should be easy (and strongly encouraged) for proposers to build funding for the currency selected by community partners.

## 5.3. Recommendations for Academic Institutions

These recommendations are directed towards academic institutions. The purpose of these recommendations is to incentivize meaningful and substantive engagement and leadership of community partners in the conception, evaluation, funding, and execution of research that impacts their lives.

### 5.3.a. Factoring Community Impact into Academic Evaluation Processes

Academic institutions should consider allowing evaluation packages to include community impact evidence documented by communities that the researcher under evaluation works with, and should develop ways to incorporate such evidence into the evaluation. This evidence might be in the form of a letter, but could also include other information that demonstrates that community partners were treated well throughout the research process, and were positively impacted by the work. Such evidence should not be in a form that overly burdens those partnering with the researcher.

There is precedent for adding evaluation criteria that are not limited to papers or grants – tenure and promotion committees often consider other artifacts like software or patents in their evaluation.

### 5.3.b. Providing Funding for Community-Based Research Projects

Academic institutions should consider mechanisms to direct internal funding towards projects that are focused on learning and building relationships with community partners. Such seed grants are effective ways to jumpstart collaborations that can be the foundation for more substantial engagements that seek external funding.

## 5.4. Recommendations Directed at Individual Researchers

These recommendations are directed towards academic researchers. The purpose of these recommendations is to provide guidance on the appropriate role of academic researchers in collaborations with community partners, and to ensure incentive alignment.

### 5.4.a. Norms for Community-Based Research

Workshop participants provided clear advice on how researchers can have meaningful and respectful interactions with community partners. These include the following:

◗ Listen first

◗ Understand that you are an outsider

◗ Follow up with the community after your research is complete–give them access to your research

◗ Act respectfully and ethically (research how to conduct yourself)

◗ Ask for consent early and often

◗ Ensure confidentiality of those you work with

These are in addition to the principles discussed in section 4:

◗ Do not expect free labor

◗ Make sure there is value alignment between stakeholders

◗ Prioritize project leadership by community members

◗ Do not treat any group like a monolith

◗ Establish common language

◗ Establish transparency of project goals and expectations

◗ Have recourse for harm if community collaborators could face any

### 5.4.b. The Role of Academic Researchers

Academics are often in positions of power, or have a seat at a table, where they are expected to speak on behalf of, or represent, a community they work with. While it is always preferable for those who are directly impacted to be able to all have a seat at the table, or at least select a community organizer/representative, those in positions of power that do understand the community should advocate for the community and share their sentiments when given the opportunity.

### 5.4.c. Expanding the Vision of What is Considered Valuable Output

Academics should think more broadly about what valuable contributions look like in order to account for positive community-based impact not just papers and quantity, which these impacts do not always lend themselves easily to.

### 5.4.d. The Tension Between Generalized and Specific Insights

Academic work is often incentivized to seek out solutions that can be generalized in all similar contexts. On the other hand, the workshop has revealed that the underlying problems and solutions – deeply rooted in humans and communities – are heterogeneous and context-specific.

Meaningful community-oriented academic research should acknowledge the variability (individual-to-individual, community-to-community, and individual-vs-community) in the sources of the problems and the ways in which different people are affected.

### 5.4.e. Research Ownership and Access

There is a natural impulse – especially nowadays – to generate research artifacts (papers, software, data) that are accessible to all. Funding agencies often require this as a condition of funding research. However, community-oriented work – as surfaced in the workshop – often demands that data collected from communities be protected, and the resulting artifacts limited in how and by whom they are used.

Community-oriented research should recognize the potentially exploitative aspects of the urge to open up access to research artifacts, and prioritize community benefits and ownership over research dissemination.

## 6. Conclusions

Collaborations between researchers in computing and the communities who are first to experience the negative impacts of AI systems can result in research directions that prevent the development of harmful AI systems, mitigate the harms of current AI systems more effectively, and build tools that prioritize human welfare. In order for these collaborations to be beneficial to community partners, we have outlined a number of recommendations. The recommendations start with broad recommendations on ensuring ethical community-based research, and then provide recommendations for specific audiences: funding agencies, academic institutions, and individual researchers.

Two particularly important recommendations are (1) the involvement of community partners in leadership positions across the life cycle of the research project, and (2) the importance of allocating funds that go directly to community-based collaborators. We hope to see more researchers and community-based partners collaborating to create a technological future that puts people's needs first.

# Workshop Participants/Workshop Report Contributors

| First Name | Last Name | Company Name |
|---|---|---|
| Noelani | Arista | McGill University |
| Ricardo | Baeza-Yates | Institute for Experiential AI, Northeastern University |
| Dameon | Brome | Above All Odds |
| Lydia X.Z. | Brown | Georgetown University |
| Tracy | Camp | Computing Research Association |
| Kade | Crockford | ACLU of Massachusetts |
| David | Danks | University of California San Diego |
| Maria | De-Arteaga | University of Texas Austin |
| Michael | Ekstrand | Boise State University |
| Sorelle | Friedler | Haverford College |
| Timnit | Gebru | Distributed AI Research Institute (DAIR) |
| Rayid | Ghani | Carnegie Mellon University |
| Sarah | Gilbert | Cornell University |
| Catherine | Gill | Computing Research Association |
| Haben | Girma | |
| Jacinta | Gonzalez | Mijente |
| Ben | Green | University of Michigan |
| Haley | Griffin | Computing Research Association |
| Alex | Hanna | Distributed AI Research Institute |
| Peter | Harsha | Computing Research Association |
| Kat | Heller | Google |
| Brian | LaMacchia | Farcaster Consulting Group |
| Daniel | Lopresti | Lehigh University and CCC |

| Keoni | Mahelona | Te Hiku Media |
|---|---|---|
| Brandeis | Marshall | DataedX Group, LLC |
| Surya | Mattu | Digital Witness Lab, Princeton University |
| Petra | Molnar | Refugee Law Lab, York University; Fellow, Harvard Law School's Berkman Klein Center |
| Melanie | Moses | University of New Mexico |
| LaTonya | Myers | Above all odds |
| Tawana | Petty | Executive Director, Petty Propolis |
| Manish | Raghavan | MIT |
| Deb | Raji | UC Berkeley |
| Fabian | Rogers | Office of NYS Senator Jabari Brisport |
| Leah | Rosenbloom | Brown University |
| Ann | Schwartz | Computing Research Association |
| Katie | Siek | Indiana University |
| Olivia | Snow | Center for Critical Internet Inquiry (C2i2), UCLA |
| Nasim | Sonboli | Brown University |
| Amos | Toh | Human Rights Watch |
| Ufuk | Topcu | The University of Texas at Austin |
| Matthew | Turk | Toyota Technological Institute at Chicago (TTIC) |
| Sepehr | Vakil | Northwestern University |
| Suresh | Venkatasubramanian | Brown University |
| Adrienne | Williams | Distributed AI Research Institute |
| Ben | Winters | Electronic Privacy Information Center (EPIC) |
| Meg | Young | Data & Society |

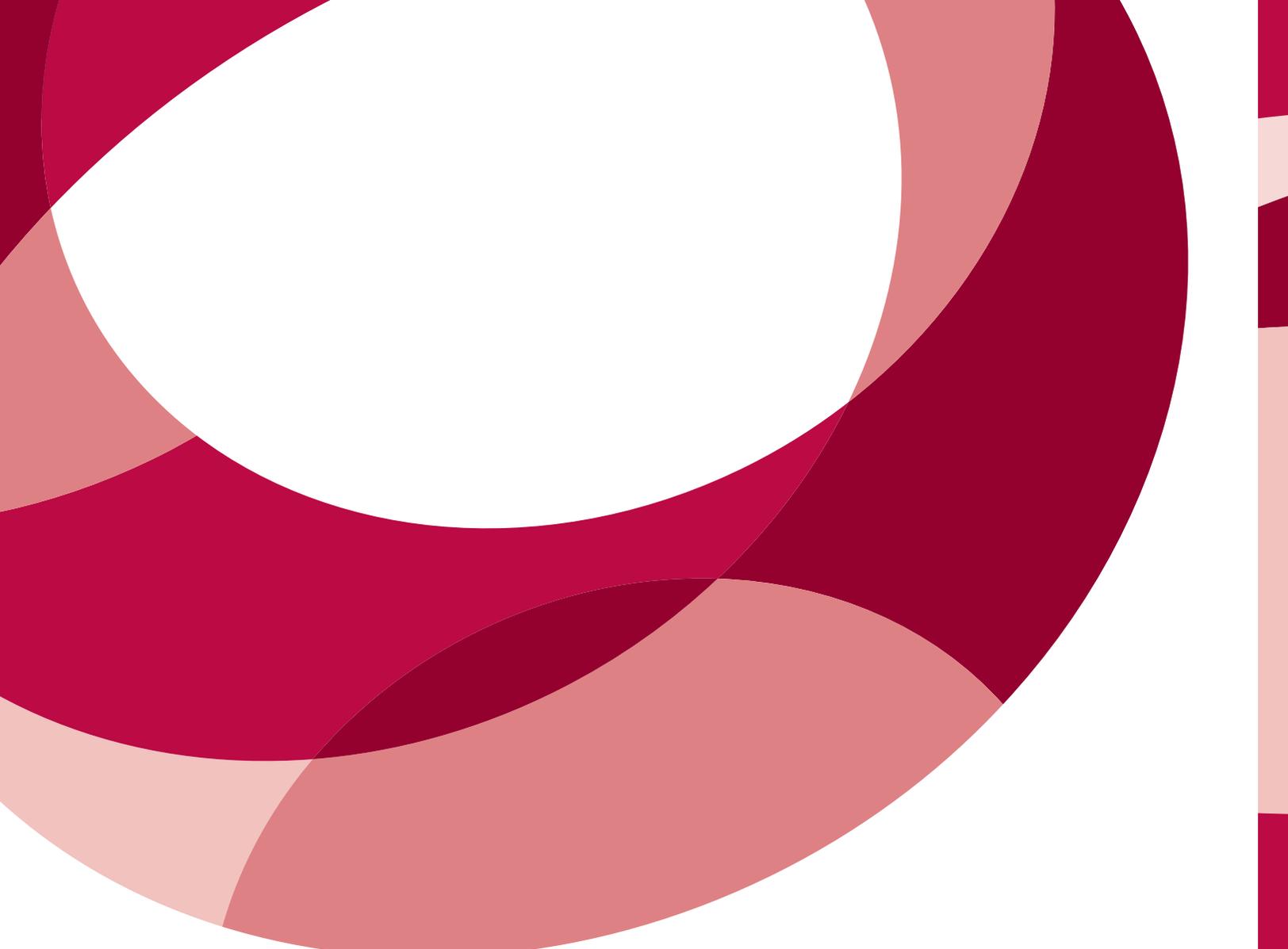

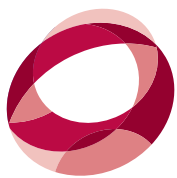 CCC
Computing Community Consortium
Catalyst

1828 L Street, NW, Suite 800
Washington, DC 20036
P: 202 234 2111  F: 202 667 1066
www.cra.org/ccc  cccinfo@cra.org